\documentstyle[prb,epsf,aps]{revtex}
\begin{document}
\draft
\twocolumn[
\hsize\textwidth\columnwidth\hsize\csname@twocolumnfalse\endcsname
\title{{\it Ab-initio} study of disorder effects on the electronic 
and magnetic structures of Sr$_2$FeMoO$_6$}

\author{T. Saha-Dasgupta \cite{snbcs}} 

\address{S. N. Bose National Center for Basic Sciences, Salt Lake, Kolkata 700 098, India } 

\author{D.D. Sarma \cite{jnc}}

\address{ Solid State and Structural Chemistry Unit, Indian Institute of Science,
Bangalore 560 012, India }

\maketitle

\begin{abstract}

We have investigated the electronic structure of ordered and disordered
Sr$_2$FeMoO$_6$ using {\it ab-initio} band structure methods. The effect
of disorder was simulated within super-cell calculations to realize several 
configurations with mis-site disorders. It is found that such disorder 
effects destroy the half-metallic ferro-magnetic state of the ordered compound.
It also leads to a substantial reduction of the magnetic moments at the 
Fe sites in the disordered configurations. Most interestingly, it is found
for the disordered configurations, 
that the magnetic coupling within the Fe sub-lattice as well as that within 
the Mo sub-lattice always remain ferro-magnetic, while the two sub-lattices couple
anti-ferromagnetically, in close analogy to the magnetic structure of the ordered
compound, but in contrast to recent suggestions. 

\end{abstract}

\vspace{0.2in}


]
\narrowtext

\section{Introduction}

Discovery of a large negative magneto-resistance, namely the colossal
magneto-resistance (CMR) \cite{DD1} in doped manganites has attracted a great deal of 
attention in recent years, because of the technological importance of
these materials being used as magnetic devices \cite{app1}. The 
pronounced influence of
the low magnetic field on the resistance of these compounds is believed
to be caused by the high degree of spin polarization of the charge
carriers, arising due to the half-metallic ferro-magnetic nature of these
materials below the magnetic transition temperature T$_c$. However the 
transition temperatures obtained in the case of the manganites 
were low for room-temperature applications, leading to a further 
search for half-metallic oxides with much higher T$_c$.
Recently Kobayashi {\it et al.} \cite{ref1} reported a large magneto-resistance 
effect with a fairly high magnetic transition temperature of
about 450 K in Sr$_{2}$FeMoO$_{6}$, a material belonging to the
class of double perovskites ($A_{2}BB'O_{6}$), where the alkaline
earth ion $A$ is Sr and transition metal ions $B$ and $B'$ are Fe and Mo
arranged in the rock-salt structure. Such a high magnetic ordering 
temperature, which is higher than that in the manganites (250 K -
350 K), is surprising and has been recently explained in terms 
of a new mechanism based on a novel magnetic interaction.\cite{prl}

In the idealized ordered structure of these double perovskite systems, the 
two transition metal ions, $B$ and $B'$, are arranged alternately along the 
cubic axes in all three directions. Specifically, in the case of Sr$_{2}$FeMoO$_{6}$
each of the two transition metal sites, namely Fe$^{3+}$ ($3d^5$, $S = 5/2$) 
and Mo$^{5+}$ ($4d^1$, $S = 1/2$) sites, are believed to be ferro-magnetically arranged within each
sub-lattice, while the two sub-lattices are supposed to be coupled anti-ferromagnetically,
giving rise to a $S = 2$ state. Recently, it has been found that it is possible to synthesize
samples of Sr$_{2}$FeMoO$_{6}$ with substantial mis-site disorders where Fe and Mo 
sites interchange their positions\cite{dd,anyother}. Such a disorder is found to have 
profound effects on the physical properties of this compound, particularly in terms of
its magnetic and CMR properties. It appears that 
such a disorder may even play a role in
determining the properties of the so-called {\it ordered} 
system \cite{unpub}, since a finite amount of 
disorder is found to be present also in such samples. 
Rietveld analysis of the X-ray powder diffraction data of Kobayashi
{\it et al.} indicated 87 $\%$ order of Fe and Mo ions. 
The saturation moment observed by Kobayashi {\it et al.} \cite{ref1} at
4.2 K was lower than the expected value of 4 $\mu_{B}$ per formula
unit considering the anti-ferromagnetic coupling between Fe$^{3+}$ and 
Mo$^{5+}$ ions, which is indicative of, as pointed out
by the authors, the mis-site type disorder of the $B$-site rock salt
arrangement. Later, the study by Sarma {\it et al.} \cite{dd} which reported
ordering at the Fe and Mo sites in their sample to be 91 $\%$, slightly 
higher than that of Kobayashi {\it et al.} showed a sharper low-field
response with improved magneto-resistive behavior.  This may be
related to the effect of disorder on the half-metallic state of the 
compound which is considered to be central to the low-field magnetic 
behavior {\it via} the spin-dependent scattering processes, as mentioned earlier.
The above studies are for bulk-sintered samples. Further studies on thin 
films of Sr$_{2}$FeMoO$_{6}$ show the saturation magnetization and the 
magneto-resistive behavior to be very different for the films grown
under different conditions \cite{asano,ref5}. The magnetization is reduced 
substantially for samples having random arrangement of Fe, Mo sub-lattices.
The reduction in the net magnetization as measured in the experiment can
occur by two distinctly different routes. One possibility is that the disorder
destroys the specific spin arrangement of the Fe and Mo
sites without any significant effect on the individual magnetic moments 
at these sites. For example, disorder may lead to an anti-ferromagnetic 
couplings between different Fe sites\cite{ref6}, instead of the 
ferro-magnetic coupling proposed for the idealized ordered structure. Alternately, 
the magnetic
moments at each individual site  may decrease due to the different chemical
environment induced by the disorder, without any effect on the 
nature of the spin ordering. The real situation may even be a combination of 
both these effects, with a simultaneous reduction in the magnetic
moments at different sites as well as a change in the nature of the
magnetic coupling between different sites.
The detailed knowledge about the disordered compound is therefore
important for better understanding of the various experimental results
and designing of samples with better magnetic and magneto-resistive 
properties. This motivated the present study, where we have employed    
the state of the art first-principles band structure method to study the
electronic structure and hence the magnetic properties of the ordered
and disordered Sr$_{2}$FeMoO$_{6}$. Our results clearly support the view that 
individual magnetic moments at each Fe site are strongly reduced due to 
the destruction of the half-metallic state in presence of disorder, while the coupling between 
the various Fe sites continue to be ferro-magnetic, in contrast to a recent suggestion\cite{ref6}.

\section{Methodology}   

Sr$_{2}$FeMoO$_{6}$ occurs in the body-centered tetragonal structure with
a space group of I4/mmm and lattice constants $a=b=5.57$ \AA~ and $c=7.90$ 
\AA~ \cite{ref1}. The oxygen atoms surrounding the Fe and Mo
sites provide the octahedral environment. The FeO$_{6}$ and MoO$_{6}$  
octahedra alternate along the three cubic axes, while Sr atoms occupy 
the hollow formed by the corners of  FeO$_{6}$ and MoO$_{6}$  
octahedra at the body-centered positions. 
In Fig. 1(a) we show the four formula unit super-cell of the
structure. To simplify the view only Fe, Mo and O atoms are shown. The
alternate positioning of transition metal Fe and Mo sites is evident
from this figure.

We have employed the linear muffin-tin orbital (LMTO) method, implemented
in the basis of screened and energy-independent muffin-tin orbitals to
study the electronic structure of the ordered and disordered 
Sr$_{2}$FeMoO$_{6}$ compound. Effect of disorder has been modeled by
super-cell calculations. 
The calculations have been performed within the framework of 
atomic sphere approximation (ASA), so that the
muffin-tin spheres were replaced by space-filling atomic spheres.
A detailed description of the LMTO-ASA formalism can be found
elsewhere \cite{ref4}. The partitioning of the space in to atom-centered
spheres was carried out ensuring that the overlaps between different atomic
spheres do not exceed 16 $ \% $. The basis set consisted of $s$, $p$, $d$, and $f$ on
Sr and Mo sites and $s$, $p$, and $d$ on Fe and O sites. The $s$ and $p$ channels on Sr
site, $d$ and $f$ channels on Mo site, $s$, $p$ and $d$ channels on Fe site and $p$ channels
on O site were considered to be active channels, while the rest were
downfolded. The scalar relativistic spin-polarized calculations with 
(6 6 6) {\bf k}-points were carried out for the self-consistency. The 
exchange-correlation part of the potential was treated using the pure 
gradient expansion of the generalized gradient
approximation (GGA). The use of GGA was found to be crucial to obtain
the correct magnetic ground-state of the compound, indicating the
possible importance of correlation effects in this compound. We have calculated the 
electronic structures of both ordered and disordered systems in order to have a 
complete comparative study within the spin-polarized approach.

%\begin{figure}
%\begin{minipage}{6.5in}
%\centerline{
%\rotatebox{0}{\resizebox{6.5in}{!}{\includegraphics{struc_or+dis.eps}}}}
%\end{minipage}
%\def\thefigure{1}
%\caption{Structure of ordered and disordered Sr$_{2}$FeMoO$_{6}$.
%Disordered structure is shown for three different
%model configurations. In the first two configurations [(b)-(c)] one pair of
%Mo and Fe have been interchanged, while in the third configuration
%[(d)] two pairs have been interchanged. White circles indicate
%Fe sites, black circles represent Mo sites while the small grey circles
%represent sites occupied by O atoms.}
%\end{figure}

In order to compare the electronic structures of the ordered and the disordered 
Sr$_{2}$FeMoO$_{6}$ on the same footing,  calculations were performed by constructing 
super-cells of four formula unit with four inequivalent Fe sites 
at (0., 0., 0.), (0., 0.5, 0.5), (0.5, 0.5, 0.) and 
(0.5, 0., 0.5) positions and four inequivalent Mo sites at 
(0., 0., 0.5), (0., 0.5, 0.0), (0.5, 0.5, 0.5) 
and (0.5, 0., 0.)  positions as shown in Fig. 1(a).
These inequivalent Fe and Mo sites are designated as Fe1, Fe2, Fe3, Fe4
and Mo1, Mo2, Mo3, Mo4, respectively. While these four sites are
crystallographically equivalent for the ordered sample, they are symmetry-distinct in presence
of mis-site disorders.
The effect of positional disorder at Fe/Mo sites is simulated 
by interchanging the Fe and Mo sites, so as to generate different chemical 
environments surrounding each inequivalent Fe and Mo sites while keeping the
concentration of Fe/Mo sites fixed. 
We have performed the disordered calculation for three different disordered
configurations, so as to check that the result is not influenced by the specific
choice of a particular disordered configuration. The first configuration is obtained by 
exchanging the Fe3 and Mo3 sites at (0.5, 0.5, 0.) and  (0.5, 0.5, 0.5) 
positions respectively, that is by interchanging one of the Fe-Mo pair 
that is nearest neighbor to each other. The second configuration is 
obtained by interchanging Fe1 and Mo3 at (0., 0., 0.) and (0.5, 0.5, 0.5), respectively;  
this amounts to an interchange of the distant Fe-Mo pairs. Fig. 1(b) and 1(c) 
show the corresponding structures with an interchange of only one Fe-Mo
pair. The structure shown in Fig. 1(b) can be derived 
from the structure shown in Fig. 1(a) by replacing the Mo site at the
body-center with an Fe atom and the face-centered Fe sites in $ab$ planes with
an Mo atom. Similarly the structure in Fig. 1(c) is derived 
by replacing the Mo site at the body-center with Fe and 
the Fe sites at the corners with
Mo. Fig. 1(d) shows the configuration that can be derived from
configuration shown in 1(b) by an additional exchange between Fe1 and
Mo1. These replacements in turn lead to distribution of 
different environment in terms of Fe and Mo sites for different 
inequivalent Fe and Mo sites.\cite{ftnt}
Table 1 summarizes the distribution of Fe and Mo
neighbors at various shells for different inequivalent Fe and Mo sites in each of
these structures. 

\section{Results and Discussion}
The spin-polarized LMTO-GGA calculation for the ordered structure 
gives the ground state to be half-metallic ferri-magnetic with a 
finite density of states in the spin-down channel and zero density 
of states in the spin-up channel and 
anti-ferromagnetic ordering between the high-spin $3d^{5}$ (S=5/2) Fe$^{3+}$
ion and the $4d^{1}$ (S=1/2) Mo$^{5+}$ ion, 
in agreement with previous results\cite{ref1,prl}. 
The magnetic moment at the Fe site was found to be +3.8 $\mu_{B}$ 
and that at the Mo site -0.3 $\mu_{B}$ in good agreement with the 
pseudo-potential calculation \cite{ref1}.
The six oxygen sites altogether contribute a magnetic moment of
about 0.5 $\mu_{B}$, thereby making the total magnetic moment of 
the system to be 4 $\mu_{B}$ per formula unit. The magnetic moment
at the oxygen sites account for the missing moments at the Fe and Mo sites
considering their Fe$^{3+}$ and Mo$^{5+}$ configurations, which
in turn indicate a strong covalency effect present in the system.

In Fig. 2(a) we show the calculated total density of states (DOS) and various
partial density of states for Sr$_2$FeMoO$_6$. The corresponding 
band-dispersions along the high-symmetry lines are shown in Fig. 2(b). 
It is to be noted that the band dispersions are plotted along the 
symmetry directions of the original primitive unit cell and not of
the super-cell. The top panels are for spin-up channel and the
bottom panels are for the spin-down channel. Since the system
orders ferri-magnetically with anti-ferromagnetic
coupling between Fe$^{3+}$ and Mo$^{5+}$ ions, the spin-up 
channel is the majority channel for Fe and minority for Mo and 
{\it vice-versa} for the spin-down channel. The predominant feature
in the band dispersion is a band-gap of about 0.8 eV 
in spin-up channel while bands cross the Fermi-level (E$_{F}$) in the spin-down channel, 
resulting in the half-metallic state.

For our discussion of density of states that is limited to an energy window 
of -8 eV to about 2 eV, we shall be primarily concerned 
with the Fe-$d$, Mo-$d$ and O-$p$ states, since the Sr derived
states appear higher in energy.  As is evident from the partial
density of states due to oxygen in the spin-up and spin-down
channels, the states spanning from -8 eV to -3.5 eV and 
separated from the rest of the states by a gap of about 0.2 eV
in the spin-up channel and about 2.5 eV in the spin-down channel are
primarily contributed by oxygen, while it has some contribution 
to states appearing close to the Fermi-energy, mixed with Fe $d$ and
Mo $d$ states. 
The presence of an approximate cubic symmetry of the octahedral
co-ordination of the oxygen atoms around 
the transition metal Fe and Mo sites, results in splitting of the $d$ levels 
of Fe and Mo sites into triply degenerate $t_{2g}$ orbitals with lower energy 
and doubly degenerate $e_{g}$ orbitals with higher energy. 
The Fe $d$ DOS, in the spin-up
channel span the energy from -3.3 eV to -2.3 eV and from -1.7 eV to
-0.2 eV, contributed by the crystal field split 
$t_{2g}$ and $e_{g}$ states respectively, giving the crystal
splitting of $\approx$ 2 eV. Fe ion being in 
the high-spin $d^{5}$ state, the 
spin-up Fe states are completely filled, lying below the Fermi-energy.
In the spin-down channel the splitting between $t_{2g}$ and $e_{g}$
states disappear due to a large band-width of Fe $t_{2g}$ states, 
which primarily comes from the hybridization with Mo
$t_{2g}$ states \cite{prl} {\it via} oxygen states. 
The $t_{2g}$ and $e_{g}$ resolved
partial density of states (not shown here) shows the exchange-splitting 
between the spin-up and spin-down Fe $t_{2g}$ and $e_{g}$ contributions to 
be about 3.5 eV.
The Mo $d$ density of states in the spin-up channel
between energy range 0.7 eV to 1.2 eV, is predominantly of $t_{2g}$
type, which is unusually narrow due to the large energy 
difference 
between the Fe and Mo $t_{2g}$ levels in the spin-up channel,
substantially suppressing the hopping processes. The crystal field split spin-up 
$e_{g}$ levels appear higher in energy, starting from 3 eV.  
Down-spin Mo $t_{2g}$, due to the hybridization with oxygen $p$ and 
Fe $t_{2g}$ in particular, as mentioned before, acquires a large
band-width spanning an energy range of 3 eV, from -1 eV to 2 eV,
thereby crossing the Fermi-level,
while the empty $e_{g}$ states as in the case of spin-up channel are
higher in energy. The exchange-splitting between the up and down Mo 
$t_{2g}$ and $e_{g}$ states is much smaller ($\approx$ 
0.5 eV) and in the opposite direction as that of Fe, in agreement
with the magnetic moment results. It should be noted here that
though the exchange splitting of the Mo 4$d$ band is much smaller
than that ($\approx$ 3.5 eV) of Fe 3$d$ band, these results
surprisingly suggest comparable Hund's coupling strengths in
the Fe $3d$ and Mo $4d$ manifolds, as suggested earlier
\cite{prl}. Since the exchange splitting of the band is given by 
the product of the coupling strength and the number of unpaired
electrons in that band, the coupling strength in Fe 3$d$ 
is in the order of 3.5/5 = 0.7 eV, where as in the Mo 4$d$, 
it is $\approx$ 0.5 eV.   

%\begin{figure}
%\begin{minipage}{6.5in}
%\centerline{
%\rotatebox{0}{\resizebox{6.2in}{!}{\includegraphics{or_d.eps}}}}
%\end{minipage}
%\def\thefigure{3}
%\caption{ (a) Total and partial density of states of ordered 
%Sr$_{2}$FeMoO$_{6}$. The solid line represents the total density of states,
%while long-dashed, dashed and dotted lines represent Mo-d, Fe-d and 
%O-p partial density of states respectively. (b) Band dispersion of 
%ordered Sr$_{2}$FeMoO$_{6}$ along the high symmetry lines.}
%\end{figure}

%\section{Electronic Structure of Dis-ordered Sr$_{2}$FeMoO$_{6}$}

Our self-consistent field calculations for the disordered models,
yield the ground state to be magnetic with a calculated magnetic
moment per formula unit reduced (3.3 $\mu_{B}$ 
for the first configuration, 3.2 $\mu_{B}$ for the second
configuration and 2.5 for the third configuration)
compared to that of the fully ordered sample (4 $\mu_{B}$), establishing a 
pronounced and systematic effect of disorder on the magnetic structure of this system; 
the calculated results are also in agreement with the experimental observation 
of a lower magnetic moment with decreasing order in such systems \cite{jpcm}.
In order to understand the microscopic origin of this reduction in
the magnetic moment, we discuss below our results in some detail.

The ground state magnetic structures for all the studied disordered
configurations were found to retain the {\it ferri-magnetic} state of the 
ordered sample in the following sense. The moments at each of the  
Fe sites were found to be ferro-magnetically aligned, irrespective of the 
near-neighbor coordinations; the same is also found to be true for
the Mo sites in every case (see Table 1). 
The two sub-lattices of Fe and Mo are found to be anti-ferromagnetically coupled, 
giving rise to the ferri-magnetic state in every disordered
configuration studied. This is in clear contrast to a previous
Monte-Carlo study \cite{ref6} of a spin
model, with local exchange interactions parameterized 
with anti-ferromagnetic coupling between the nearest neighbor transition 
metal sites, irrespective of the specific occupancies at these sites. 
Not surprisingly, this model suggests anti-ferromagnetically coupled 
Fe-O-Fe bonds, whenever two Fe ions form near-neighbor pairs. On
the other hand, our results clearly show that the Fe-O-Fe interaction
is actually {\it ferro-magnetic}, 
{\it e.g.} for the first configuration, the Fe2, Fe3, Fe4 sites  
which are co-ordinated by Fe sites in the nearest neighbor cation shell,
as shown in Table 1. This holds true even
for the case of Fe1 site in the second configuration, where all the six
nearest neighboring cations along the $x$-, $y$- and $z$-directions 
are occupied by Fe. 
Thus, the experimentally observed reduction of the magnetic moment
in disordered samples does not
arise from an anti-ferromagnetic coupling of Fe-O-Fe 
bonds in a disordered sample; instead 
the reduction in the net magnetic moment is
caused by the reduction of individual magnetic moments at different
inequivalent Fe sites depending upon the various distribution of the 
number of Fe and Mo sites in the neighbor shells. 
Our full-fledged {\it ab-initio}
calculations therefore establish a very different behavior than that
obtained from the previous model calculation, where the exchange
couplings were {\it assumed} to be inherently anti-ferromagnetic 
for the nearest neighbor Fe-O-Fe arrangements.

As Table I shows for the disordered configurations, the number of Fe and Mo 
atoms in various neighbor shells differ from that of the ordered arrangement.
As expected the deviation is largest for Fe and Mo sites that are   
interchanged in position. We notice that the magnetic moments at the Fe
sites are strongly influenced by the number of the Mo atoms in the
nearest neighbor shell. More the number of 
Mo neighbors surrounding each Fe atom, larger is the magnetic
moment. When the Fe site is surrounded by only Mo sites in the
nearest cation shell the magnetic moment at the Fe site 
in the disordered configuration is 3.8 $\mu_{B}$,
same as that of the ordered calculation.
Reducing the number of
Mo near-neighbors by 2 from that of ordered arrangement reduces the
magnetic moment to 3.1 $\mu_{B}$, 
as can be seen for Fe2 and Fe4 in configuration 1 and Fe2, Fe3 and Fe4 in 
configuration 2 given in Table 1. Reducing the number of Mo coordinating 
a central Fe 
even further decreases 
the magnetic moment at Fe sites down to 2.0-2.5
$\mu_{B}$. The reduction of the magnetic moment
at the Fe sites with the reduction in number of Mo neighbors may be 
explained by considering the fact that the charge transfer energy 
between the Fe and Mo is substantial. As a result the presence of Mo 
neighbors decreases the hopping probability from the Fe site, enhancing 
the importance of electron correlation and stabilizing a larger
magnetic moment. On the contrary, the
reduction in number of Mo neighbors and increase in number of
Fe neighbors has an opposite effect, the hopping to neighboring Fe sites
enhances the bandwidth, making the central Fe site less correlated
with a reduced magnetic moment.

%\begin{figure}
%\begin{minipage}[b]{6.3in}
%\centerline{
%\rotatebox{0}{\resizebox{6.2in}{!}{\includegraphics{dis_d.eps}}}}
%\end{minipage}
%\def\thefigure{3}
%\caption{Total and partial density of states of disordered 
%Sr$_{2}$FeMoO$_{6}$ for three different configurations. Density of
%states corresponding to first, second and third configuration are
%shown in 3(a), 3(b) and 3(c). As before,
%the solid line represents the total density of states,
%while long-dashed, dashed and dotted lines represent Mo-d, Fe-d and 
%O-p partial density of states respectively. Upper panels correspond
%to up spin and down panels correspond to down spin. The zero of the
%energy marked by the vertical line has been set at Fermi-energy}
%{\bf [TSD: Figures too small ? But I can not think of any better arrangement.]}
%\end{figure}

In Figs. 3(a) and (b), we show the DOS and the partial Fe $d$, Mo $d$ and 
O $p$ DOS for the disordered configurations 
1 and 3; the results from configuration 2 are very similar to 
those from configuration 1. The most evident consequence 
of the mis-site disorder that is common in all the results is a destruction 
of the half-metallic ferri-magnetic state due to the collapse of the gap present 
between the Fe $e_g$ and Mo $t_{2g}$ spin-up states in the 
ordered sample. Since the disorder has been introduced between Fe
and Mo sub-lattices, the high-lying Sr-derived states at energy 2
eV and above remains more or less unaffected. The energy states
lying low in energy, from about 
-9 eV to -4 eV in the spin-up channel and from -9 eV to -2 eV
in the spin-down channel, are dominated by oxygen $p$ states while 
Fe $d$ and Mo $d$ contribute to states close to E$_{F}$ in
close similarity with the gross features of the ordered DOS.
The smearing effect of the disorder
reduces the gap between oxygen $p$ derived bands and the rest of
the bands leading to significant mixing between the oxygen $p$
dominated states and the states having contribution from Fe
$t_{2g}$ states in spin-up channel appearing in energy
approximately between -4 eV to -2 eV, while
the gap between the oxygen derived states and the hybrid states
arising from admixture of Fe $t_{2g}$, Mo $t_{2g}$ and oxygen $p$ crossing
the Fermi-level in spin-down channel reduces to about 0.8 eV in 
configuration 1 and collapses in configuration 3, compared to 
a gap of about 2.5 eV in the case of the ordered sample (see
Fig. 2). In the spin-up channel, disordering
fills up the gap between crystal field split Fe $t_{2g}$ and Fe $e_{g}$
states observed in ordered DOS which are still below the Fermi-energy as well
as the gap between the filled Fe $d$ states and the sharp density of states
arising due to narrow Mo $t_{2g}$ bands lying above E$_{F}$.
We notice an overall broadening effect due to the disorder with
the density of states reduced compared to that of the ordered
calculation, arising from a spread in the site diagonal energies induced by 
different Madelung potentials at the crystallographically
inequivalent sites. The electronic structure in the disordered
configurations are also affected by the off-diagonal disorder
arising from changes in the cation neighbors and a consequent
lowering of the symmetry. It is important to note that there is
only a moderate polarization of the states at and near E$_{F}$ in
all the disordered configurations, in sharp contrast to the 100$\%$
polarization in the ordered sample. Since the transport mechanism
involves charge carriers within a small energy range of E$_{F}$,
the absence of a significant spin-polarization of the states near
E$_{F}$ should have a pronounced effect in the 
low-field CMR properties of
such disordered systems. It is interesting to note that
experimentally there is indeed a complete suppression of the 
remarkable low-field CMR effect in the disordered samples \cite{dd}.
 
It should be kept in mind that the super-cell
calculations are actually ordered calculations that make use of
the translational symmetry, which is not the case of a truly disordered
material. Furthermore, even though we have used reasonably large
super-cells with four formula units containing 40 atoms, it is still
a small number and therefore the various plausible configurations
that we can probe are limited. 
In particular, if a system experimentally approaches a chemically
phase separated scenario with only Fe ions somewhere spatially
separated from the region with only Mo ions, such situations cannot 
be described within the present size of the super-cell. 
It is possible that experimentally such phase separated scenarios
are actually achieved in certain extreme cases.
We have probed the limit where disorder is still microscopic and
does not lead to a larger length-scale phase separation \cite{EPL}.
In this context it will be 
ideal to perform a truly disordered calculation that is capable of probing local
environment effects \cite{ASR}. 
 
In conclusion, we 
have performed electronic structure calculations for ordered and disordered
Sr$_{2}$FeMoO$_{6}$ with positional disorder at the Fe/Mo sites. 
The effect of disorder has been modeled by super-cell calculations.
Disordering is found to destroy the half-metallic nature of the ordered 
compound; moreover, disorder leads to a significant reduction 
in the net magnetic moment. The reduction in the 
magnetic moment is caused by the decrease of the individual magnetic
moments of different inequivalent Fe 
sites due to the various distributions of the
neighbors and consequent band-width/correlation 
effects. Contrary to the result of a recent Monte Carlo study \cite{ref6},
even the near-neighbor 
Fe-Fe and Mo-Mo interactions continue to be ferro-magnetic
also in the disordered configurations 
with an anti-ferromagnetic coupling between the 
Fe and Mo sub-lattices.

\newpage
\onecolumn
\begin{small}
\begin{table}
\begin{tabular}{||c|c|c|c|c|c|c|c||}\hline
& Sites & dist=3.94 $\AA$ & dist=3.95 $\AA$ & dist=5.57 $\AA$ & dist=5.58 
$\AA$ & Mag. & Net.Mom.\\
& & & & & & Mom. & per f.u. \\
& & [ 1/2\{ 0 1 0\} ] & [1/2\{ 0 0 1\} ] & [1/2 \{1 1 0 \}] &
[ 1/2\{1 0 1\}] & $\mu_{B}$ & $\mu_B$ \\\hline
Ordered&Fe&4 Mo&2 Mo&4 Fe&8 Fe&3.8& 4.0\\
& Mo & 4 Fe & 2 Fe & 4 Mo & 8 Mo & -0.3 & \\\hline
& Fe 1 & 4 Mo & 2 Mo & 4 Mo & 8 Fe & 3.8 & \\
& Fe 2 & 2 Fe + 2 Mo & 2 Mo & 4 Fe & 4 Fe + 4 Mo & 3.1 & \\
& Fe 3 & 4 Fe & 2 Mo & 4 Mo & 8 Mo & 2.0 & 3.1 \\
Disordered & Fe 4 & 2 Fe + 2 Mo & 2 Mo & 4 Fe & 4 Fe + 4 Mo & 3.1 &
\\
(Config. 1)& Mo 1 & 4 Fe & 2 Fe & 4 Fe & 8 Mo & -0.3 & \\
& Mo 2 & 2 Mo + 2 Fe & 2 Fe & 4 Mo & 4 Mo + 4 Fe
& -0.3 & \\
& Mo 3 & 4 Mo & 2 Fe & 4 Fe & 8 Fe & -0.2 & \\
& Mo 4 & 2 Mo + 2 Fe & 2 Fe & 4 Mo & 4 Mo + 4 Fe & -0.3 & \\\hline 
& Fe 1 & 4 Fe & 2 Fe & 4 Mo & 8 Mo & 2.5 & \\
& Fe 2 & 2 Fe + 2 Mo & 2 Mo & 4 Fe & 4 Fe + 4 Mo & 3.1 & \\
& Fe 3 & 4 Mo & 2 Fe & 4 Mo & 8 Fe & 3.1 & \\
Disordered & Fe 4 & 2 Fe + 2 Mo & 2 Mo & 4 Fe & 4 Fe + 4 Mo & 3.1 &
\\
(Config. 2)& Mo 1 & 4 Fe & 2 Mo & 4 Fe & 8 Mo & -0.3 & 3.2 \\
& Mo 2 & 2 Fe + 2 Mo & 2 Fe & 4 Mo & 4 Mo + 4 Fe
& -0.3 & \\
& Mo 3 & 4 Mo & 2 Mo & 4 Fe & 8 Fe & -0.1 & \\
& Mo 4 & 2 Mo + 2 Fe & 2 Fe & 4 Mo & 4 Mo + 4 Fe & -0.3 & \\ \hline
Disordered & Fe  & 4 Fe & 2 Mo & 4 Fe & 8 Mo & 2.5 & \\
(Config. 3)& Mo  & 4 Mo & 2 Fe & 4 Mo & 8 Fe & -0.3 & 2.5 \\ \hline
\end{tabular}
\caption{Co-ordination of Mo and Fe sites surrounding each inequivalent
Fe and Mo sites for different neighbor shells for the ordered and three
disordered configurations. The shells are marked with distances and the 
vectors connecting the sites. Due to the slight
deviation from the cubic symmetry the vectors involving translations
along $c-$axis are having slightly different distances as those involving
translations along $a-$ and $b-$axis. The last two columns give the magnetic
moment at various inequivalent sites and the net magnetic moment.}
\end{table}
\end{small}
\newpage
{\large{\bf Figure Captions}} \\ \\
\noindent
1. Structure of ordered and disordered Sr$_{2}$FeMoO$_{6}$.
Disordered structure is shown for three different
model configurations. In the first two configurations [(b)-(c)] one pair of
Mo and Fe have been interchanged, while in the third configuration
[(d)] two pairs have been interchanged. Big grey and black circles indicate
Fe and Mo sites while the small grey circles represent sites
occupied by O atoms. \\ \\
\noindent
2.(a) Total and partial density of states of ordered 
Sr$_{2}$FeMoO$_{6}$. The solid line represents the total density of states,
while long-dashed, dashed and dotted lines represent Fe $d$, Mo $d$ and 
O $p$ partial density of states respectively. (b) Band dispersion of 
ordered Sr$_{2}$FeMoO$_{6}$ along the high symmetry lines.  \\ \\
\noindent
3.Total and partial density of states of disordered 
Sr$_{2}$FeMoO$_{6}$. Density of
states corresponding to first and third configuration are
shown in 3(a) and 3(b). As before,
the solid line represents the total density of states,
while long-dashed, dashed and dotted lines represent Fe $d$, Mo $d$ and 
O $p$ partial density of states respectively. Top panels correspond
to spin-up and bottom panels correspond to spin-down channels. 
The zero of the energy marked by the vertical line has been set at 
Fermi-energy. 


\begin{thebibliography}{99}
\bibitem[12]{ftnt} The redistribution process makes the different Fe
(Mo) sites to be
equivalent for the third configuration with the same local
environment.
\end{thebibliography}
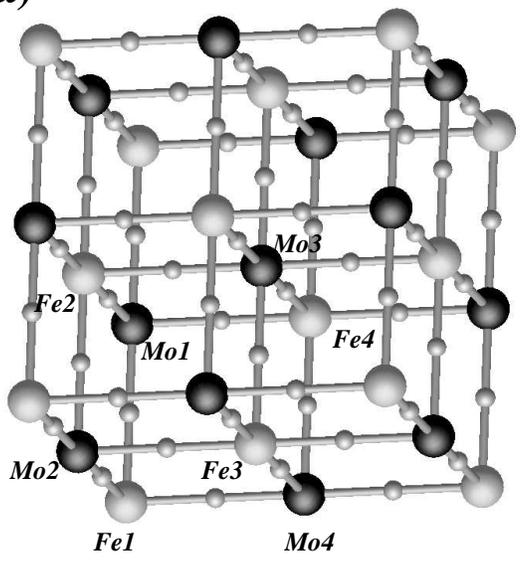
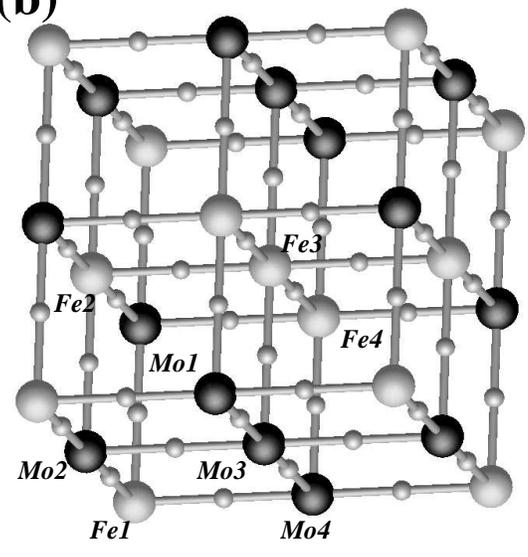
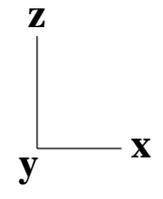
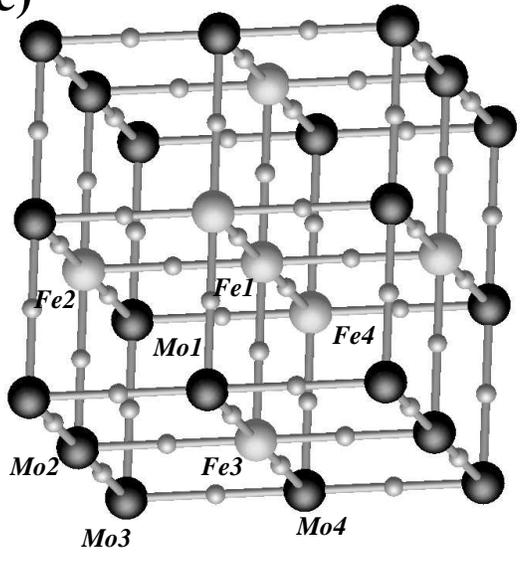
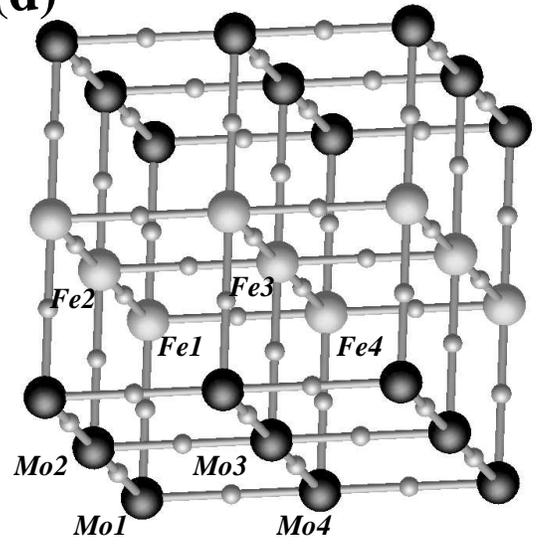
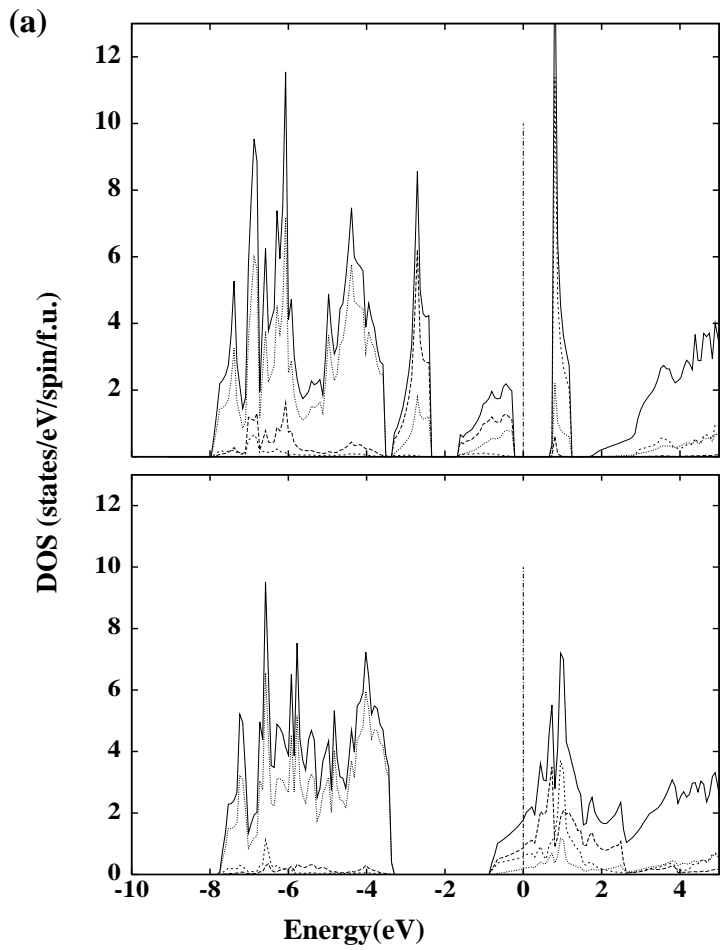
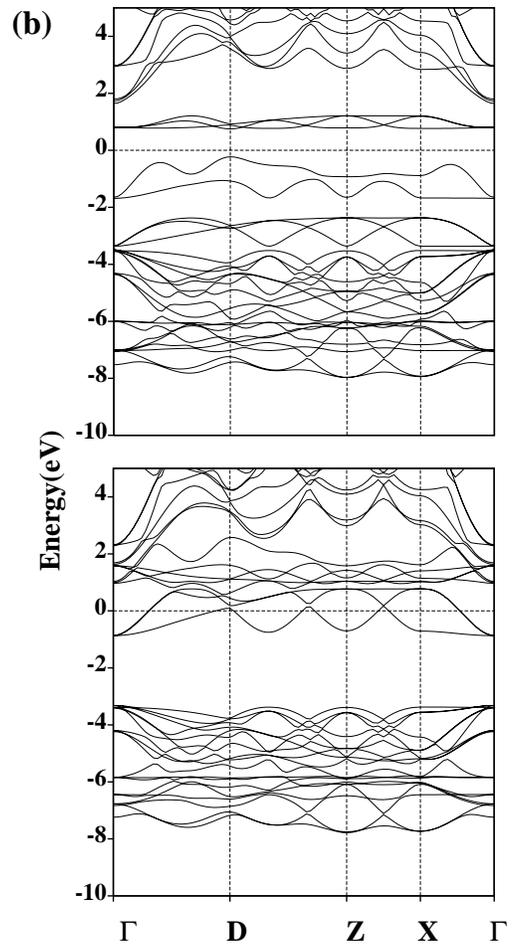
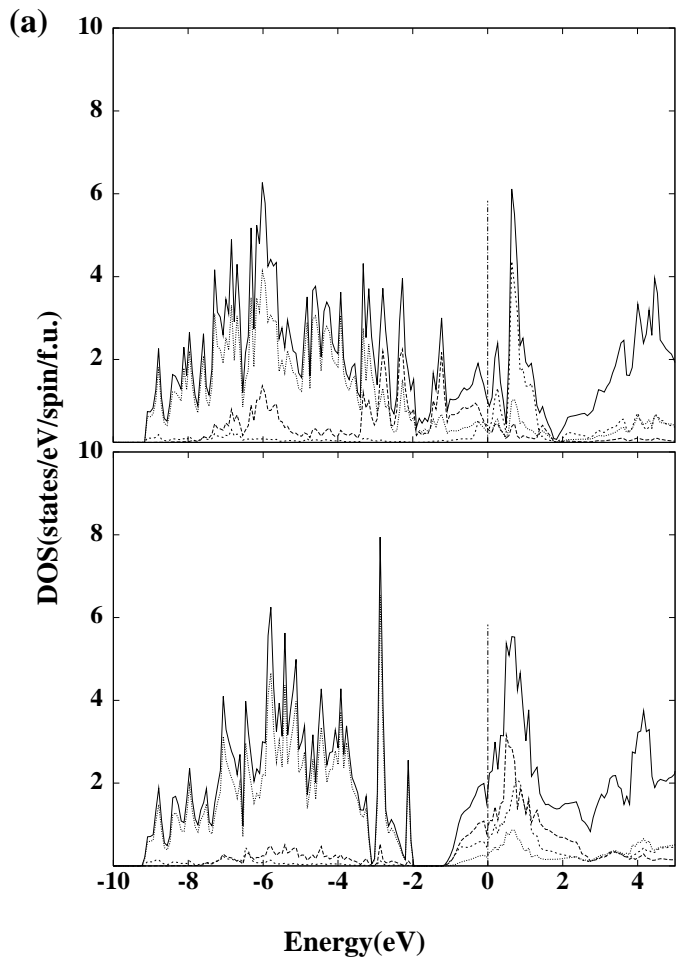
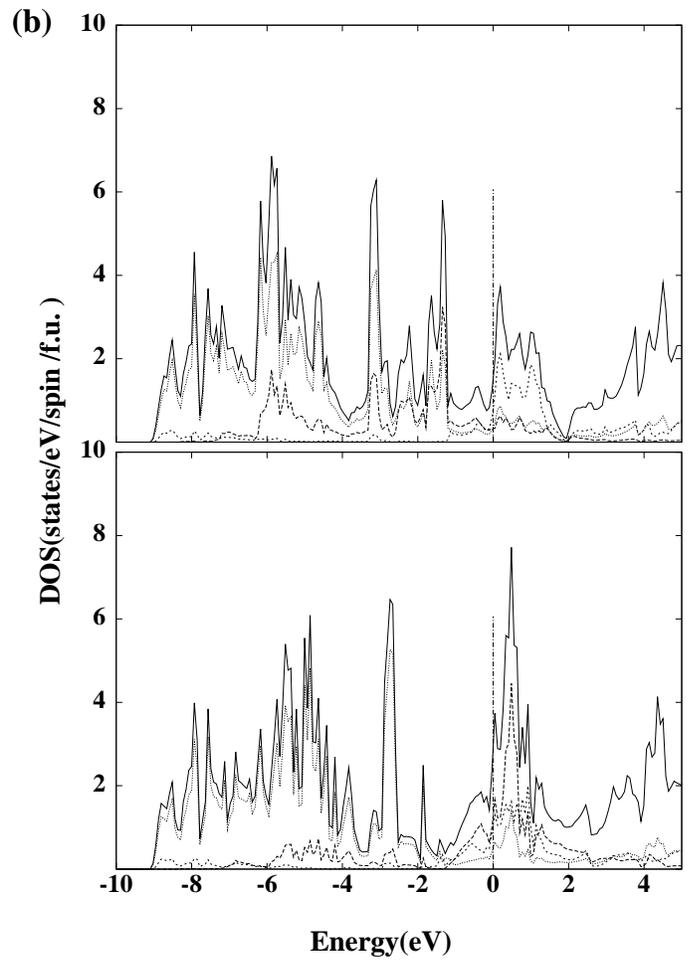
\end{document}